\documentclass[aps,prl,reprint]{revtex4-2}
\usepackage{blindtext}
\usepackage{graphics}
\usepackage{hyperref}
\usepackage{amsmath}
\usepackage{physics}
\usepackage{graphicx}
\usepackage[export]{adjustbox}
\usepackage[capitalise]{cleveref}
\usepackage[font=small]{caption}
\graphicspath{ {./images/} }
\DeclareGraphicsExtensions{.png,.eps}
\begin{document}
\newcommand{\Br}{{\bf r}}
\newcommand{\BR}{{\bf R}}
\newcommand{\BK}{{\bf K}}
\title{Deterministic vortices evolving from partially coherent fields}
\author{Wenrui Miao$^{1}$}
\author{Yongtao Zhang$^{2}$}
\email{yongtaozhang@mnnu.edu.cn}
\author{Greg Gbur$^{1,}$}
\email{gjgbur@uncc.edu}
\affiliation{$^{1}$Department of Physics and Optical Science, UNC Charlotte, Charlotte, North Carolina 28223, USA}
\affiliation{$^{2}$College of Physics and Information Engineering, Minnan Normal University, Zhangzhou 363000, China}

\begin{abstract}
It has long been assumed that there is an intrinsic conflict between optical vortices and partial coherence, in that deterministic phase vortices do not appear in partially coherent fields. We demonstrate, however, that it is possible to construct a beam that has no deterministic vortices in the source plane yet evolves a deterministic vortex at a specified propagation distance.
\end{abstract}

\maketitle
Over the past few decades, research into optical wavefield singularities \cite{mssmvv:pio,mrdkohmjp:pio,gjg:so:2017} has become ubiquitous and has led to many interesting physical insights and potential applications, including imaging \cite{jhlgfegjgas:prl:2006} and free-space optical communications \cite{gibson2004free,jwyyyimfnayyhhyryysdmtaew:np:2012}. The most common singularity is an optical vortex, a line of zero intensity in a three-dimensional wavefield around which the phase has a circulating or helical structure. The handedness of the phase structure imbues a vortex beam with orbital angular momentum, and such a beam can be used to trap or rotate particles \cite{simpson1997mechanical}, or create light-driven micromachines \cite{kldgg:oe:2004}.

Optical coherence theory is another field of optics that been researched intensely in recent years, with many potential applications \cite{KOROTKOVA202043}. Partially coherent beams can be structured to have unusual propagation characteristics like self-focusing \cite{Ding:17}, and it is now well-known that partially coherent beams are in many circumstances resistant to the distortions caused by atmospheric turbulence \cite{Gbur:14}. These observations have motivated the study of partially coherent beams to improve free-space optical communications \cite{oklcarlp:oe:2004}.

The possibilities of optical vortices and partial coherence make it natural to ask whether there is some benefit to combining their effects.  However, it has long been assumed that there is an intrinsic conflict between vortices and partial coherence: a vortex arises as a deterministic phase structure, while the phase is random in a partially coherent field. Though optical vortices can appear in two-point coherence functions, it was shown that field vortices associated with a zero of intensity are not typically found in a partially coherent field \cite{gbur2004hidden}.  A partially coherent field that has a vortex imparted upon it in the source plane, through use of a spiral phase plate for example, would appear to always lose that deterministic vortex on propagation.

However, it was recently found that a certain class of partially coherent beams, now called Rankine vortex beams, will evolve a deterministic vortex as they propagate into the far zone \cite{zhang2020partially}.  This arises because a Rankine vortex beam is the Fourier transform of a Gaussian Schell-model vortex beam, with a deterministic vortex at its core, and so the evolution of a Rankine beam into the far zone results in a deterministic vortex embedded in a partially coherent field. 

Of most interest for applications, however, is a partially coherent beam that can be designed to have a deterministic vortex appear at any desired propagation distance. In this letter, we demonstrate that a partially coherent beam can be designed, through the use of fractional Fourier transforms (FracFTs), to manifest a deterministic vortex at any range and for any degree of spatial coherence in the source plane; we refer to such beams as deterministic vortex beams (DVBs).  Our analytic results are illustrated with examples, and show that the relationship between phase singularities and partial coherence is more complicated than generally believed.

To characterize the behavior of partially coherent wavefields, we work in the frequency domain and use the cross-spectral density (CSD) function $W(\Br_1,\Br_2,\omega)$, which may be formally defined as \cite{wolf1982new}
\begin{equation}
W(\mathbf{r}_1,\mathbf{r}_2,\omega)=\langle U^*(\mathbf{r}_1,\omega)U(\mathbf{r}_2,\omega) \rangle_\omega
\label{eq1}
\end{equation}
where $\langle \dots \rangle_\omega$ represents the average over an ensemble of monochromatic fields and the asterisk denotes the complex conjugate.  For a quasi-monochromatic field, the CSD at central frequency $\omega$ provides an excellent description of the overall behavior of the field, and we suppress $\omega$ as a functional argument for brevity moving forward.

To design a field that creates deterministic vortices at a specified distance, we start with a Gaussian Schell-model vortex (GSMV) beam of the form,
\begin{equation}
\begin{split}
W_{0}(\Br_1,\Br_2)=&\frac{(x_1-iy_1)(x_2+iy_2)}{\sigma^2}e^{-r_1^2/2\sigma^2}\\
& e^{-r_2^2/2\sigma^2}e^{-\vert\Br_2-\Br_1\vert^2/2\delta^2 },
\end{split}
\label{eq2}
\end{equation}
where $\sigma$ represents the waist width of the beam and $\delta$ represents the transverse spatial correlation length. The terms of the form $x+iy$ represent a deterministic vortex, which appears at the origin and has a zero of the spectral density $S(\Br) = W(\Br,\Br)$. This function is of Schell-model form because the spectral degree of coherence depends only upon the difference variable, i.e.
\begin{equation}
    \mu_0(\Br_2-\Br_1) = e^{-\vert\Br_2-\Br_1\vert^2/2\delta^2 }.
\label{eq3}
\end{equation}

To create a DVB, we take two-dimensional FracFTs with respect to the variables $\Br_1$ and $\Br_2$. The most elegant approach to do this is to first write the Schell-model degree of coherence of the beam in terms of its Fourier transform,
\begin{equation}
\mu_0(\BR) = \int \tilde{\mu}_0(\BK)e^{i\BK\cdot\BR}d^2K,
\label{eq4}
\end{equation}
where
\begin{equation}
\tilde{\mu}_0(\textbf{K})=\frac{1}{(2\pi)^2}\int e^{-R^2/2\delta^2}e^{-i\textbf{K}\cdot{\textbf{R}}}d^2R = \frac{\delta^2}{\pi}e^{-K^2\delta^2/2}. 
\label{eq5}
\end{equation}
The cross-spectral density may then be expressed in the form,
\begin{equation}
W_{0}(\Br_1,\Br_2) = \int \tilde{\mu}_0(\BK) U_0^\ast(\Br_1,\BK)U_0(\Br_2,\BK)d^2K, \label{W0:tilde}
\end{equation}
where $U_0(\Br,\BK)$ represents a monochromatic tilted vortex beam,
\begin{equation}
U_0(\Br,\BK)=\frac{(x+iy)}{\sigma}e^{-r^2/2\sigma^2}e^{i\textbf{K}\cdot{\textbf{r}}}.
\label{eq7}
\end{equation}
 
A FracFT is a generalization of the classical Fourier transform with an order parameter $\alpha$. The transform is defined such that
$\alpha=0$ represents the identity operation and $\alpha=\pi/2$ represents the ordinary Fourier transform operation. The FracFT can be implemented optically using a single lens system, as proposed by Lohmann \cite{lohmann1993image}.

The 2-D FracFT can be expressed as an integral transform \cite{vn:imajam:1980},
\begin{equation}
 U_\alpha(\textbf{r},\BK)= \int_{-\infty}^{\infty} \textbf{F}_\alpha(\textbf{r},\textbf{r}')U_0(\textbf{r}',\BK) \,d^2r', 
 \label{eq8}
\end{equation}
where $\textbf{F}_\alpha(\textbf{r},\textbf{r}')$ represents the 2-D FracFT kernel defined as
\begin{equation}
\textbf{F}_\alpha(\textbf{r},\textbf{r}')=\frac{ie^{-i\alpha}}{2\pi\sigma^2\sin\alpha}e^\frac{-i\cot\alpha r^2}{2\sigma^2}e^{\frac{i\textbf{r}\cdot\textbf{r}'}{\sigma^2\sin\alpha}}e^{-\frac{i\cot\alpha r'^2}{2\sigma^2}}.
\label{eq9}
\end{equation}
The FracFT is typically defined as a transform over dimensionless variables; to implement it in transforming a beam, we had to choose a length scale for the transform. In Eq.~(\ref{eq9}), this scale is taken as $\sigma$ so that the beam width of the FracFT tilted beams is independent of the choice of the FracFT order $\alpha$ in the source plane.

After applying the FracFT to the tilted beams in the source plane, Fresnel diffraction can be used to propagate them to any desired distance. The field distribution along propagation is expressed as
\begin{equation}
U_\alpha(\textbf{r},\BK,z)=\int\!\int \textbf{G}(\textbf{r},\textbf{r}^\prime) \textbf{F}_\alpha(\textbf{r}^\prime,\textbf{r}^{\prime\prime}) U_0(\textbf{r}^{\prime\prime},\BK)d^2r''d^2r',
\label{eq10}
\end{equation}
Where $\textbf{G}(\textbf{r},\textbf{r}^\prime)$ is the Fresnel diffraction kernel, given by
\begin{equation}
\textbf{G}(\textbf{r},\textbf{r}^\prime)=\frac{e^{ikz}}{i\lambda z} e^{\frac{ik\lvert \textbf{r}-\textbf{r}^\prime\rvert^2}{2z}}.
\label{eq11}
\end{equation}
Evaluating the integrals of Eq.~(\ref{eq10}), we find the fractional Fourier field propagated to a distance $z$ has the form,
\begin{equation}
\begin{split}
U_\alpha(\textbf{r},\BK,z)
=&\frac{-e^{ikz}e^{-i\alpha}}{4\beta^4A^2\sigma}e^{\frac{-(\sin{\alpha}+i\cos{\alpha})}{4\beta^2A\sigma^2}r^2}e^{\frac{-\textbf{K}\cdot\textbf{r}}{2\beta^2A}}\\
&e^{\frac{-K^2}{4A}}[(x+K_x \beta^2)+i(y+K_y \beta^2)].
\end{split}
\label{eq12}
\end{equation}
In this expression, 
\begin{equation}
\beta^2 \equiv \sigma^2\sin\alpha - \frac{z}{k}\cos\alpha,
\label{eq13}
\end{equation}
and 
\begin{align}
A&\equiv\frac{i\Tilde{\beta}^2}{2\beta^2\sigma^2}+\frac{1}{2\sigma^2}, \\
\Tilde{\beta}^2&\equiv\sigma^2\cos{\alpha}+\frac{z}{k}\sin{\alpha}.
\label{eq15}
\end{align}

It is to be noted that each tilted beam has a vortex at the shifted position $\beta^2\BK$; however, if $\beta^2=0$, every tilted beam will have its vortex at the origin, and the cross-spectral density will therefore also have a deterministic vortex at the origin. Solving $\beta^2=0$ for $z$, we find that we recover the Schell-model vortex beam at distance
\begin{equation}
    z_0=k\sigma^2\tan\alpha.\label{z0:def}
\end{equation}
Because the tangent function takes on all positive values from zero to infinity as $\alpha$ ranges from $0$ to $\pi/2$,  we may choose our source parameters to produce a deterministic vortex at any desired propagation distance.

We may determine the cross-spectral density by a formula analogous to Eq.~(\ref{W0:tilde}), 
\begin{equation}
W_\alpha(\textbf{r}_1,\textbf{r}_2,z)=\int \tilde{\mu}_0(\textbf{K})U^*_\alpha(\textbf{r}_1,\BK,z)U_\alpha(\textbf{r}_2,\BK,z)\, d^2K.
\label{eq17}
\end{equation}
Substituting from Eq.~(\ref{eq5}) and Eq.~(\ref{eq12}) into the above integral yields
\begin{equation}
\begin{split}
W_\alpha(\textbf{r}_1,\textbf{r}_2,z)&=\frac{\delta^2}{16\beta^8\lvert A \rvert^4 \sigma^2}e^{-\frac{(\sin{\alpha}+i\cos{\alpha})}{4\beta^2A\sigma^2}r_2^2}\\
&e^{-\frac{(\sin{\alpha}-i\cos{\alpha})}{4\beta^2A^*\sigma^2}r_1^2} e^{\frac{(\textbf{r}_1A+\textbf{r}_2A^*)^2}{16\beta^4\lvert A \rvert^4 \eta}}\\
&\Biggl[\frac{r_1^2A^2+r_2^2A^{*2}+2\lvert A\rvert^2(x_1x_2+y_1y_2)}{16\lvert A\rvert^4\eta^3}\\
-&\frac{r_1^2A+r_2^2A^{*}+(A+A^{*})(x_1-iy_1)(x_2+iy_2)}{4\lvert A\rvert^2\eta^2}\\
&+\frac{(x_1-iy_1)(x_2+iy_2)}{\eta} +\frac{\beta^4}{\eta^2}\Biggr],
\end{split}
\label{eq18}
\end{equation}
where $\eta\equiv\frac{\delta^2}{2}+\frac{1}{4A}+\frac{1}{4A^*}$. A detailed derivation process of the CSD along propagation is included in the Supplemental Material. The phase of this cross-spectral density will manifest a deterministic vortex at the distance $z_0$.

\begin{figure}[htb]%
\includegraphics[width=8cm]{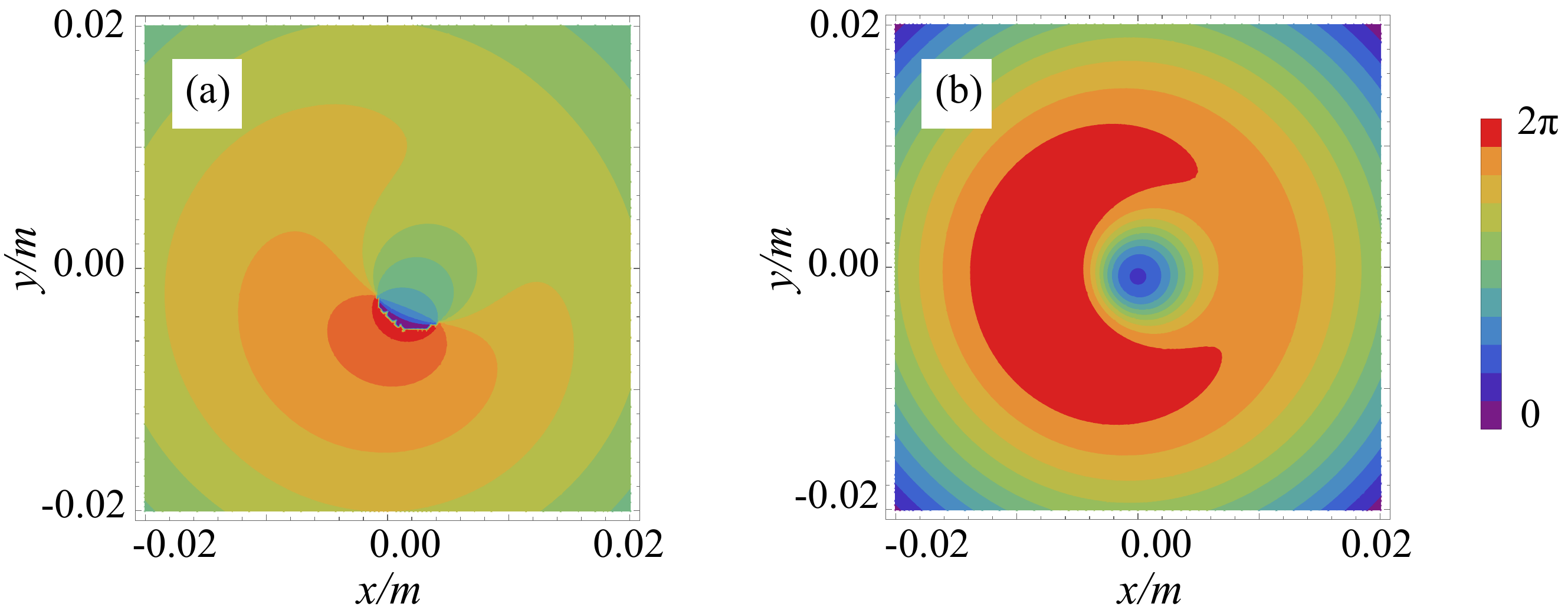}%
\caption{Cross-spectral density of a DVB in the source plane. Here $\alpha=0.308$, $\sigma=5\times10^{-3}$m, $\delta=0.01$m. The observation point $\Br_1$ is located at (a) ($0.3$mm,$0.3$mm), (b) ($0.1$mm,$0$mm).}
\label{fig1}%
\end{figure}

For the following examples, the beam waist width is set to $5$ mm and the wave number $k=\frac{2\pi}{5}\times10^7$ m. Figure \ref{fig1} shows the phase of the CSD in the source plane $z=0$ for different values of the reference point $\Br_1$. It can be seen that the positions of the vortices, if they even exist, depend upon the position of the observation point, and are therefore singularities of the correlation function and not deterministic vortices.

\begin{figure}[htb]%
\includegraphics[width=8cm]{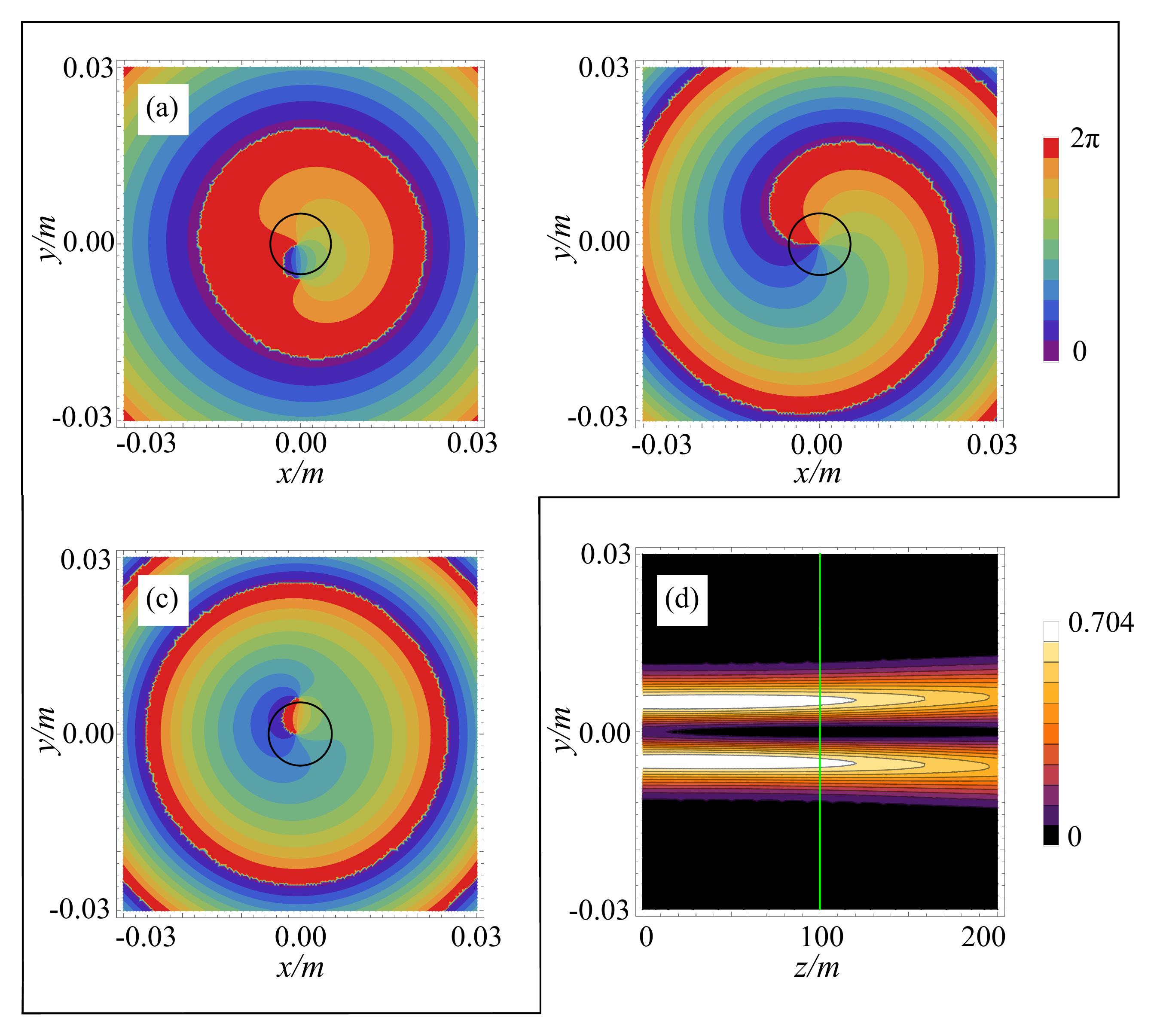}%
\caption{Phase of a DVB beam with $z_0=100$m. (a) z=$77$m, (b) z=$100$m, (c) z=$123$m; (d) intensity distribution along propagation. Here, in (a)-(c),the observation point $r_1$ is located at ($0.1$mm,$0$mm), $\alpha=0.308$, $\sigma=5\times10^{-3}$m, $\delta=0.01$m. The black circle represents the radius $\sigma$; in (d) the green line indicates the position where the deterministic vortex is located.}
\label{fig2}%
\end{figure}

Figure \ref{fig2} shows the phase of the cross-spectral density and the spectral density in the vicinity of the special distance, here chosen to be $z_0=100$ m, with $\alpha=0.308$.  Figures \ref{fig2}(a)-(c) show the phase distribution at different propagation distances; the circle indicates the radius $\sigma$ of the beam width for scale.    It can be seen that there is a single vortex at the special distance $z_0$, and as one moves away from the special distance the CSD has a pair of opposite-handed vortices in the correlation function. Figure \ref{fig2}(d) shows the cross-section of the spectral density along propagation; the green line indicates the position $z_0$.

These deterministic vortices of the DVBs only appear over a finite range around the distance $z_0$. We may estimate this ``depth of field'' of deterministic vortices as the range over which the single deterministic vortex is the only one that appears within a circle with the radius of the beam width.  In \cref{fig2}, it can be seen that from $z=77$ m to $z=123$ m there is always only one phase singularity in the beam center within the beam width, making the depth of field approximately $46$m for this case. Within this depth of field, the position of the vortex does not depend significantly on the position $\Br_1$ of the reference point; outside this depth of field, the vortex position depends strongly on $\Br_1$.

\begin{figure}[htb]%
\includegraphics[width=8cm]{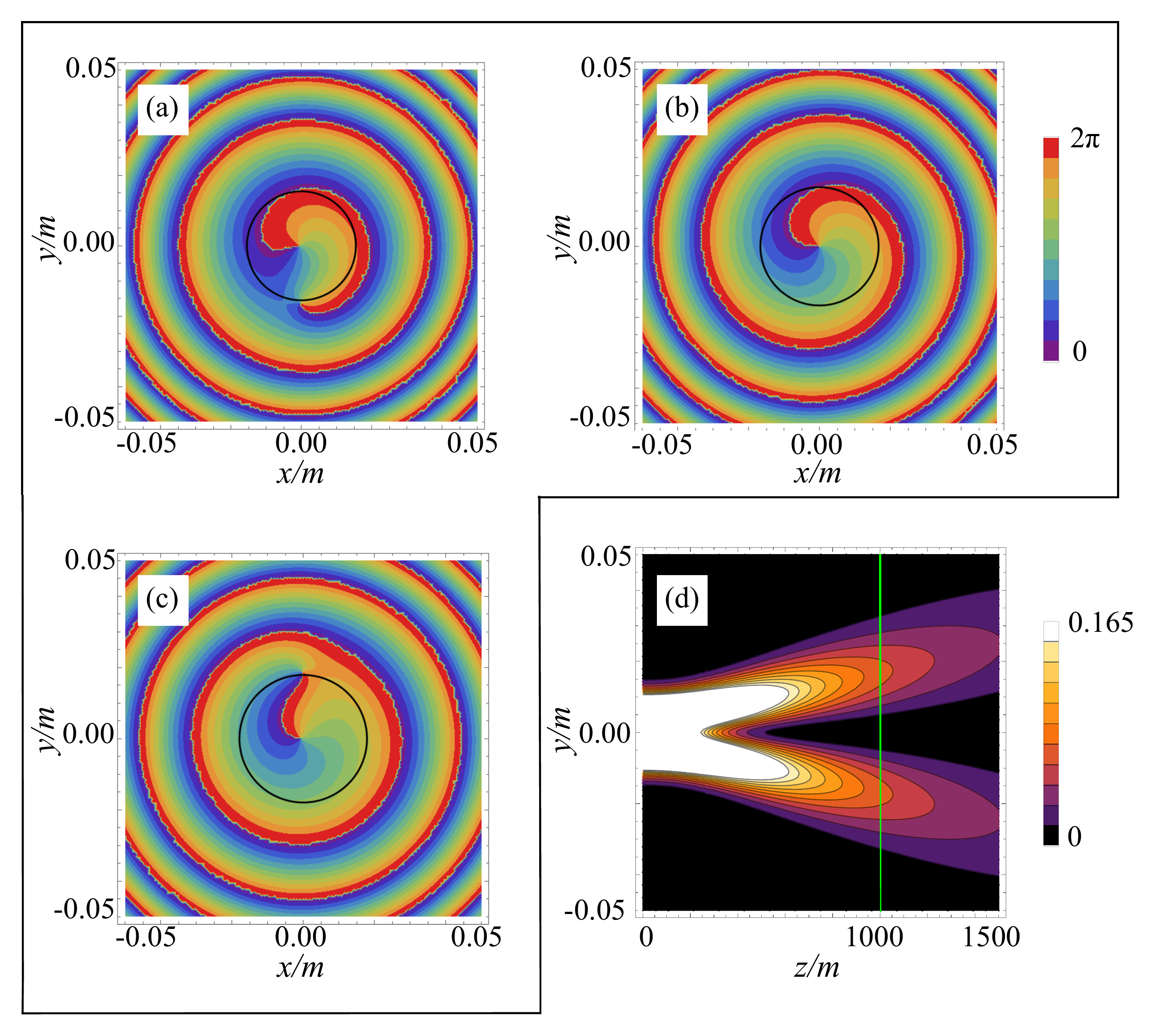}%
\caption{Creating a Gaussian Schell-model vortex beam at $z_0=1000$m. (a) z=$920$m, (b) z=$1000$m, (c) z=$1080$m; (d) intensity distribution along propagation. Here, $\alpha=1.2664$, $\sigma=5\times10^{-3}$m, $\delta=0.01$m. }
\label{fig3}%
\end{figure}

As noted above, these deterministic vortices can be placed at any propagation distance, even quite far away. Figure \ref{fig3} shows a DVB for the choice $z_0=1000$ m; in this case $\alpha=1.2664$. The correlation width is still $\delta=0.01$ m, so that the depth of field is approximately $160$m, which is around the same order of magnitude as in \cref{fig2}. This ability to maintain a deterministic vortex in a partially coherent field over long propagation distances has potential application in free-space optical communication and long-range remote sensing.

\begin{figure}[htb]%
\includegraphics[width=8cm]{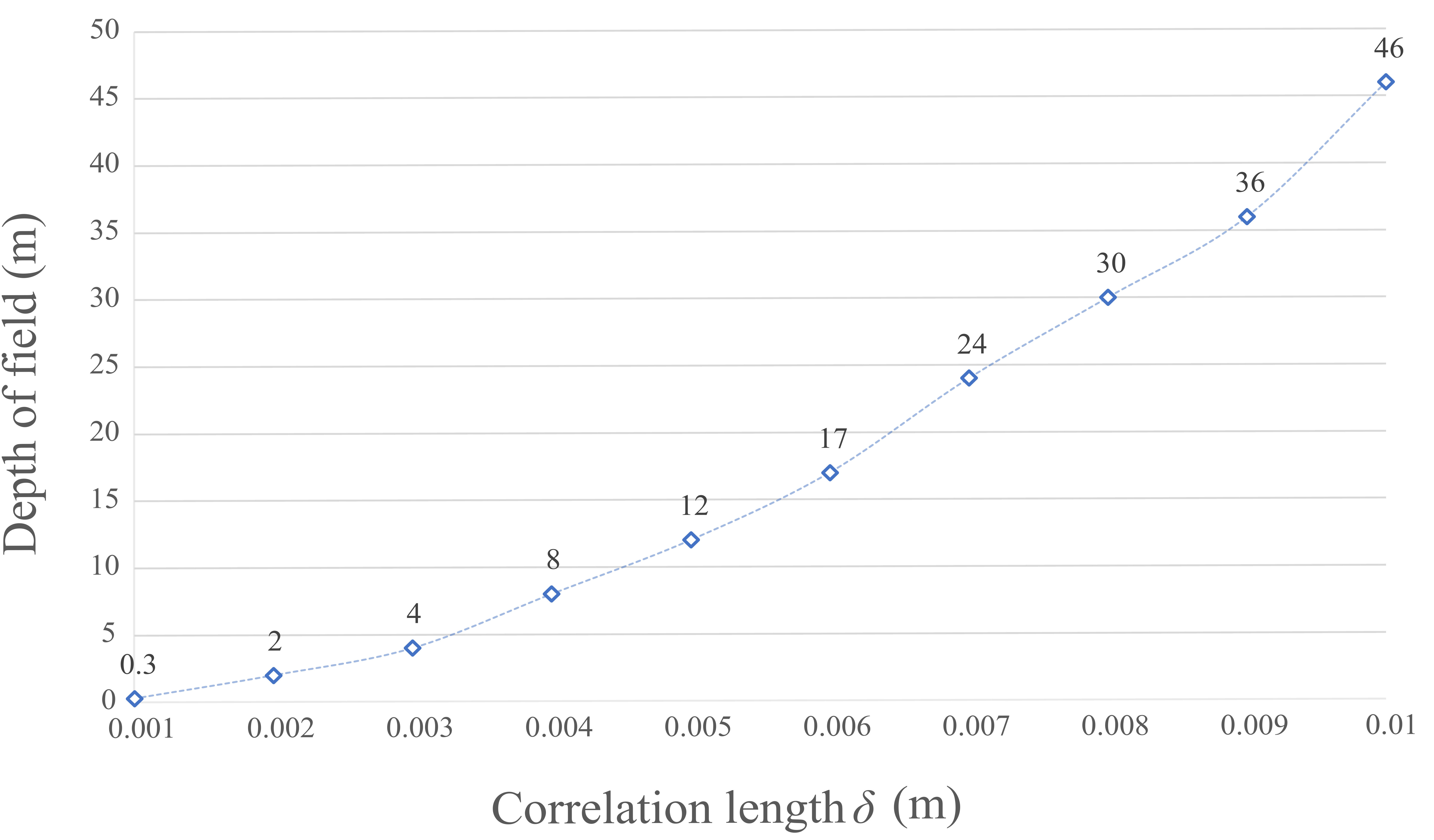}%
\caption{Depth of field of the deterministic vortex as correlation length $\delta$ increases from $0.001$m to $0.01$m. Here, the observation point $r_1$ is located at ($0.1$mm,$0$mm), $\alpha=0.308$, $\sigma=5\times10^{-3}$m. }
\label{fig4}%
\end{figure}

The correlation width $\delta$ does not affect the position of deterministic vortices, as can be seen by Eq.~(\ref{z0:def}). However, it does change the depth of field. Figure \ref{fig4} shows the calculated depth of field over a range of values of the correlation length $\delta$. It can be seen that the depth of field decreases dramatically as the spatial coherence decreases. 

\begin{figure}[htb]%
\includegraphics[width=8cm]{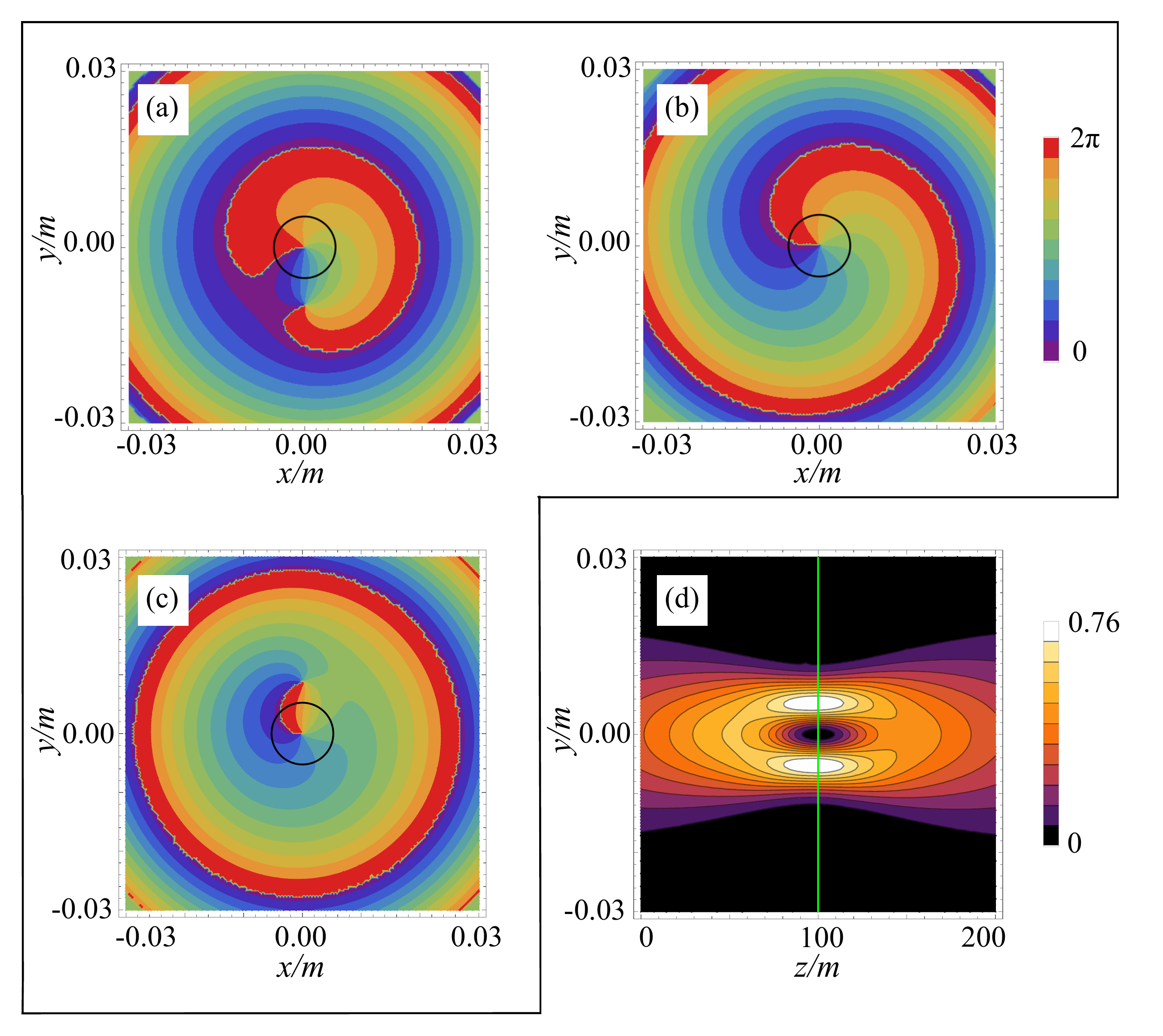}%
\caption{Decrease the transverse correlation length $\delta$ to $0.001$m. (a) z=$99.8$m, (b) z=$100$m, (c) z=$100.1$m; (d) intensity distribution along propagation. Here, $\alpha=0.308,\sigma=5\times10^{-3}$m.}
\label{fig5}%
\end{figure}

Figure \ref{fig5} shows the case $z_0=100$ m but $\delta = 0.001$ m. From the selected phase images, it can be seen that the depth of field is reduced to about $0.3$ m.  Lower spatial coherence therefore corresponds with a smaller depth of field. Furthermore, it can be seen from \cref{fig5}(d) that the field exhibits minor self-focusing in this low coherence limit.

This self-focusing effect can be traced to the effect of the FracFT on the constituent components of the beam. Equation (\ref{eq12}) indicates that, even for $z=0$, the effect of the FracFT is to tilt the beam in one direction and spatially shift it in the opposite direction. This naturally creates a converging beam. For beams with low spatial coherence, and contributions from large $\BK$ values, this convergence is stronger than the natural diffractive spreading of the beam.

In summary, we have demonstrated that it is possible to design partially coherent beams which manifest a deterministic vortex at any desired propagation distance in free space. In this letter, we only considered a vortex beam of order 1, but deterministic vortices of any order can in principle be created by changing the order of the GSMV beam at the source.

Though FracFTs are often implemented optically using a lens system, it is important to note that such an optical system is not necessary to realize the beams described here.  The only thing that is required is that the field matches the cross-spectral density of Eq.~(\ref{eq18}) in the source plane. This can be done, for instance, by encoding a spatial light modulator with an appropriate statistical ensemble. 

Our results may find use in the synthesis of partial coherence and optical vortices for free-space optical communication. They also demonstrate that the study of the relationship between coherence and singular optics can still yield unexpected effects.\\

 Greg Gbur is supported by the Office of Naval Research (N00014-20-1-2558) and the Air Force Office of Scientific Research (FA9550-21-1-0171). Yongtao Zhang is supported by the National Natural Science Foundation of China 12174173, the Fujian Provincial Natural Science Foundation of China 2022J02047.

\bibliographystyle{apsrev4-2} 
\bibliography{fracFT} 
\end{document}


\newcommand{\Br}{{\bf r}}
\newcommand{\BR}{{\bf R}}
\newcommand{\BK}{{\bf K}}
\title{Supplemental Material for“Deterministic vortices evolving from partially coherent fields"}
\author{Wenrui Miao$^{1}$, Yongtao Zhang$^{2}$, Greg Gbur$^{1,*}$}

\address{$^{1}$Department of Physics and Optical Science, UNC Charlotte, Charlotte, North Carolina 28223, USA}
\address{$^{2}$College of Physics and Information Engineering, Minnan Normal University, Zhangzhou 363000, China}

\begin{abstract}

\end{abstract}

\maketitle
In this supplemental material, we show a detailed derivation process of the cross-spectral density (CSD) along propagation, which is Eq.~(18) in the original manuscript.

We begin with a Gaussian Schell-model vortex (GSMV) beam

\begin{equation}
\begin{split}
W_{0}(\textbf{r}_1,\textbf{r}_2)=&(x_1-iy_1)(x_2+iy_2)e^{-r_1^2/2\sigma^2}\\
& e^{-r_2^2/2\sigma^2}e^{-\vert\textbf{r}_2-\textbf{r}_1\vert^2/2\delta^2 }.
\end{split}
\label{eq1}
\end{equation}
First, we write the spectral degree of coherence in terms of its Fourier transform, 
\begin{equation}
\mu_0(\BR) = \int \tilde{\mu}_0(\BK)e^{i\BK\cdot\BR}d^2K,
\label{eq2}
\end{equation}
where
\begin{equation}
\begin{split}
\tilde{\mu}_0(\textbf{K})=&\frac{1}{(2\pi)^2}\int e^{-R^2/2\delta^2}e^{-i\textbf{K}\cdot{\textbf{R}}}d^2R, \\
=&\frac{\delta^2}{\pi}e^{-K^2\delta^2/2}. \\
\end{split}
\label{eq3}
\end{equation}
The cross-spectral density may then be expressed in the form,
\begin{equation}
W_{0}(\Br_1,\Br_2) = \int \tilde{\mu}_0(\BK) U_0^\ast(\Br_1,\BK)U_0(\Br_2,\BK)d^2K, 
\label{eq4}
\end{equation}
where $U_0(\Br,\BK)$ represents a monochromatic tilted vortex beam,
\begin{equation}
U_0(\Br,\BK)=\frac{(x+iy)}{\sigma}e^{-r^2/2\sigma^2}e^{i\textbf{K}\cdot{\textbf{r}}}.
\label{eq5}
\end{equation}
The 2-D FracFT for a tiled vortex beam can be defined as an integral transform
\begin{equation}
 U_\alpha(\textbf{r},\BK)= \int_{-\infty}^{\infty} \textbf{F}_\alpha(\textbf{r},\textbf{r}')U_0(\textbf{r}',\BK) \,d^2r', 
 \label{eq6}
\end{equation}
where $\textbf{F}_\alpha(\textbf{r},\textbf{r}')$ represents the 2-D FracFT kernel defined as
\begin{equation}
K_\alpha(\textbf{r},\textbf{r} ')=\frac{ie^{-i\alpha}}{2\pi\tau^2\sin\alpha}e^\frac{-i\cot\alpha r^2}{2\tau^2}e^{\frac{i\textbf{r}\cdot\textbf{r}'}{\tau^2\sin\alpha}}e^{-\frac{i\cot\alpha r'^2}{2\tau^2}}.
\label{eq7}
\end{equation}

Proper $\tau$ value needs to be chosen so that the beam width is invariant regardless of the choice of the FracFT order $\alpha$ in the source plane. To find the beam width, we integrate all the exponential terms of a normally incident Gaussian beam over the source plane,
\begin{equation}
\begin{split}
I=&\int e^{\frac{-i\cot\alpha r^2}{2\tau^2}}e^{\frac{i\Br\Br'}{\tau^2\sin\alpha}}e^{\frac{-i\cot\alpha r'^2}{2\tau^2}}e^{\frac{-r'^2}{2\sigma^2}}d^2r' \\
=&e^{\frac{-i\cot\alpha r^2}{2\tau^2}}\int e^{-\left(\frac{i\cot\alpha}{2\tau^2}+\frac{1}{2\sigma^2}\right)r'^2}e^{\frac{i\Br\Br'}{\tau^2\sin\alpha}}d^2r'.
\end{split}
\label{eq8}
\end{equation}
We use the relation,
\begin{equation}
e^{-A(r-B)^2}=e^{-Ax^2+2ABx'-AB^2},
\label{eq9}
\end{equation}
For the above integral, 
\begin{equation}
A=\frac{i\cot\alpha}{2\tau^2}+\frac{1}{2\sigma^2}, \ B=\frac{ir}{i\cos\alpha+\frac{\tau^2}{\sigma^2}\sin\alpha},
\label{eq10}
\end{equation}
and thus,
\begin{equation}
\begin{split}
I=& e^{AB^2}\int e^{-A(\Br\prime-B)^2}d^2r'= \frac{\pi}{A}\frac{\pi}{\frac{i\cot\alpha}{2\tau^2}+\frac{1}{2\sigma^2}}e^{AB^2},
\end{split}
\label{eq11}
\end{equation}

where
\begin{equation}
e^{AB^2}=e^{-\frac{\sigma^2}{2\tau^2\sin{\alpha}(i\cos\alpha\sigma^2+\tau^2\sin{\alpha})}r^2}.    
\label{eq12}
\end{equation}
The beam width $\omega$ is found as 
\begin{equation}
\begin{split}
 \omega=& \Re  \left[-\frac{\sigma^2}{2\tau^2\sin{\alpha}(i\cos\alpha\sigma^2+\tau^2\sin{\alpha})}\right]^\frac{-1}{2}\\&=\frac{\sqrt{2}(\cos^2\alpha\sigma^4+\tau^4\sin^2\alpha)^\frac{1}{2}}{\sigma}.
\end{split}
\label{eq13}
\end{equation}
Beam width goes from $\sqrt{2}\sigma$ to $\frac{\sqrt{2}\tau^2}{\sigma}$, as the FracFT order $\alpha$ increases from $0$ to $\pi/2$. So in order to keep all fractional beams sharing the same width, $\tau$ needs to be set as $\sigma$.\\
Then, the 2-D FracFT kernel is expressed as
\begin{equation}
\textbf{F}_\alpha(\textbf{r},\textbf{r}')=\frac{ie^{-i\alpha}}{2\pi\sigma^2\sin\alpha}e^\frac{-i\cot\alpha r^2}{2\sigma^2}e^{\frac{i\textbf{r}\cdot\textbf{r}'}{\sigma^2\sin\alpha}}e^{-\frac{i\cot\alpha r'^2}{2\sigma^2}}.
\label{eq14}
\end{equation}

After applying the FracFT to the tilted beams in the source plane, Fresnel diffraction can be used to propagate them to any desired distance. The field distribution along propagation is expressed as
\begin{equation}
U_\alpha(\textbf{r},\BK,z)=\int\textbf{G}(\textbf{r},\textbf{r}^\prime) U_\alpha(\textbf{r}^{\prime},\BK)d^2r',
\label{eq15}
\end{equation}
Where $\textbf{G}(\textbf{r},\textbf{r}^\prime)$ is the Fresnel diffraction kernel, given by
\begin{equation}
\textbf{G}(\textbf{r},\textbf{r}^\prime)=\frac{e^{ikz}}{i\lambda z} e^{\frac{ik\lvert \textbf{r}-\textbf{r}^\prime\rvert^2}{2z}}.
\label{eq16}
\end{equation}
The 2-D FracFT and Fresnel diffraction integrals can be combined to write as
\begin{equation}
U_\alpha(\textbf{r},\BK,z)=\int\int \textbf{G}(\textbf{r},\textbf{r}^\prime) \textbf{F}_\alpha(\textbf{r}^\prime,\textbf{r}^{\prime\prime}) U_0(\textbf{r}^{\prime\prime},\BK)d^2r''d^2r',
\label{eq17}
\end{equation}
The integral over $r'$ can be calculated first to combine the FracFT and Fresnel kernel into the combined kernel

\begin{equation}
\begin{split}
\textbf{H}(\textbf{r},\textbf{r}^{\prime\prime})&=\int \textbf{G}(\textbf{r},\textbf{r}^\prime) \textbf{K}_\alpha(\textbf{r}^\prime,\textbf{r}^{\prime\prime}) d^2r'   \\
&=\frac{ie^{ikz}e^{-i\alpha}}{2\pi \beta^2}e^{\frac{ik}{2z}r^2}e^{\frac{-i\cot{\alpha}\textbf{r}^{\prime\prime2}}{2\sigma^2}} e^{-\frac{i\gamma(\textbf{r}^{\prime\prime}-\textbf{r}/\gamma)^2}{2\beta^2}},
\end{split}
\label{eq18}
\end{equation}
where $\beta^2\equiv\sigma^2\sin{\alpha}-\frac{z}{k}\cos{\alpha}$, 
 $\gamma\equiv\frac{z}{k\sigma^2\sin{\alpha}}$.

Then the field distribution along
propagation is obtained by the following integral
\begin{equation}
U_\alpha(\textbf{r},\BK,z)=\int \textbf{H}(\textbf{r},\textbf{r}^{\prime\prime}) U_0(\textbf{r}^{\prime\prime},\BK)d^2r''.
\label{eq19}
\end{equation}
Substituting Eq.~(\ref{eq5}) and Eq.~(\ref{eq18}) and after lengthy calculations, Eq.~(\ref{eq19}) yields
\begin{equation}
\begin{split}
U_\alpha(\textbf{r},\BK,z)=&\frac{-e^{ikz}e^{-i\alpha}}{4\beta^4A^2\sigma}e^{\frac{-(\sin{\alpha}+i\cos{\alpha})}{4\beta^2A\sigma^2}r^2}e^{\frac{-\textbf{K}\cdot\textbf{r}}{2\beta^2A}}\\
&e^{\frac{-K^2}{4A}}[(x+k_x \beta^2)+i(y+K_y \beta^2)],
\end{split}
\label{eq20}
\end{equation}
where
$A\equiv\frac{i\Tilde{\beta}^2}{2\beta^2\sigma^2}+\frac{1}{2\sigma^2}$, $\Tilde{\beta}^2\equiv\sigma^2\cos{\alpha}+\frac{z}{k}\sin{\alpha}$.

Then, the cross-spectral density along propagation can be obtained using a formula analogous to Eq.~(\ref{eq4}),
\begin{equation}
W_\alpha(\textbf{r}_1,\textbf{r}_2,z)=\int \tilde{\mu}_0(\textbf{K})U^*_\alpha(\textbf{r}_1,\BK,z)U_\alpha(\textbf{r}_2,\BK,z)\, d^2K.
\label{eq21}
\end{equation}
Substituting from Eq. (\ref{eq3}) and Eq. (\ref{eq20}) into the above
integral yields
\begin{equation}
\begin{split}
W_\alpha(\textbf{r}_1,\textbf{r}_2,z)&=\frac{\delta^2}{16\beta^8\lvert A \rvert^4 \sigma^2}e^{-\frac{(\sin{\alpha}+i\cos{\alpha})}{4\beta^2A\sigma^2}r_2^2}\\
&e^{-\frac{(\sin{\alpha}-i\cos{\alpha})}{4\beta^2A^*\sigma^2}r_1^2}e^{\frac{(\textbf{r}_1A+\textbf{r}_2A^*)^2}{16\beta^4\lvert A \rvert^4 \eta}}\\
&\Biggl[\frac{r_1^2A^2+r_2^2A^{*2}+2\lvert A\rvert^2(x_1x_2+y_1y_2)}{16\lvert A\rvert^4\eta^3}\\
-&\frac{r_1^2A+r_2^2A^{*}+(A+A^{*})(x_1-iy_1)(x_2+iy_2)}{4\lvert A\rvert^2\eta^2}\\
&+\frac{(x_1-iy_1)(x_2+iy_2)}{\eta}+\frac{\beta^4}{\eta^2}\Biggr],
\end{split}
\label{eq22}
\end{equation}
where $\eta\equiv\frac{\delta^2}{2}+\frac{1}{4A}+\frac{1}{4A^*}$.

At the special distance $z_0$, the CSD reduces to
\begin{equation}
\begin{split}
W_\alpha(\textbf{r}_1,\textbf{r}_2,z_0)&=\frac{\delta^2}{16\beta^8\lvert A \rvert^4 \sigma^2\eta}
e^{\frac{(\textbf{r}_2-\textbf{r}_1)^2}{16\beta^4A^2 \eta}}e^{-\frac{(\sin{\alpha}+i\cos{\alpha})}{4\beta^2A\sigma^2}r_2^2}\\
&e^{-\frac{(\sin{\alpha}-i\cos{\alpha})}{4\beta^2A^*\sigma^2}r_1^2}
(x_1-iy_1)(x_2+iy_2),
\end{split}
\label{eq23}
\end{equation}
which is in the form of a GSMV beam.